\documentclass{article}
\usepackage{amssymb,epsfig}
\begin{document}
\title{Self--similar and charged spheres \\in the diffusion approximation}
\author{W. Barreto and A. Da Silva\\
Laboratorio de F\'isica Te\'orica,
Departamento de F\'isica,\\
Escuela de Ciencias, N\'ucleo de Sucre,\\
Universidad de Oriente,
Cuman\'a, Venezuela.}
\maketitle
\begin{abstract}
We study spherical, charged and self--similar distributions of
matter in the diffusion approximation.
We propose a simple, dynamic 
but physically meaningful solution. For such a solution we obtain a model in which
the distribution becomes static and changes to dust.
The collapse is halted with damped mass oscillations about the 
absolute value of the total charge. 
\end{abstract}

\section{Introduction}
Several authors have considered charged distributions of matter 
\cite{b65,f69,s71,b71,b75,cd78,hp85,o90,lz91,o91,fms95},
although it is well known that astrophysical objects are not significantly
charged. Nevertheless, in some stages of the gravitational
collapse even a small amount of charge can change the final state
of the body. Some interesting features of charged collapsing matter
justify any effort to get physical insight by studying this problem.
 For instance, naked singularities can
be prevented \cite{p93}; the final geometrical structure left over
after the complete collapse of a spherically symmetric charged source
and of a chargeless rotating star are similar \cite{l95}; Cauchy horizons,
gravitational repulsion and perhaps traversable wormholes are
also possible \cite{o90}.

If the mathematical treatment is simplified,
the evolution of a charged distribution of matter can be followed by
the Einstein--Maxwell equations.
In this paper we explore the self--similar gravitational collapse of
charged spheres in the diffusion approximation. It is well known that
dissipation due to the emission of massless particles is a characteristic
process in the evolution of massive stars. The only plausible mechanism
to carry away the bulk of binding energy of the collapsing star, leading
to a neutron star or black hole, is neutrino emission \cite{ks79}.
It seems clear that the
free--streaming process is associated with the initial stages of
the collapse, while
the diffusion approximation becomes valid toward the final stages.
The junction conditions at the boundary of a charged and dissipative
sphere have been considered with outgoing heat flow and radiation flux
\cite{ds87,bd89}.
On the other hand, the field equations 
admit homothetic motion
\cite{ct71,op90,hp91,lz90}.
Applications of homothetic simila\-rity range from modeling
black holes to producing counterexamples to the cosmic censorship
conjecture \cite{ch74,bh78a,bh78b,eimm86,cy90,l92,b95,cc98}.
In particular, homothetic charged and isotropic (or anisotropic)
 fluids have been
 studied \cite{ps79,t84,hp85}.

We observe in the literature that much work has been done 
under static conditions and dusty charged matter. Also authors
make additional assumptions such as equations of state or
relationships between metric variables \cite{t84,hm84}. In this paper 
we obtain a dynamical model from a simple solution to
the homothetic motion, without any additional supposition except
spherical symmetry and self--similarity. 
In section
2 we write the field equations and the junction conditions.
The equations at the surface of the distribution of matter are
presented in section 3. In section 4 we write the symmetry
equations to describe self--similarity. In section 5, we show
an example from a simple solution which we discuss in the last section.

\section{Field equations and matching}
Let us consider a non--static distribution of matter which is spherically
symmetric and consists of charged fluid of energy density $\rho$, pressure
 $p$, electric charge density $\sigma$ and radiation energy flux $q$
diffusing in the radial direction, as measured by a local Minkowskian
observer comoving with the fluid. In radiation coordinates \cite{b64} the
metric takes the form
\begin{equation}
\label{im}ds^2=e^{2\beta}\bigg(\frac Vrdu^2+2du\,dr\bigg)
-r^2\bigg(d\theta ^2+\sin\theta^2\,d\phi ^2\bigg), 
\end{equation}
where $\beta $ and $V$ are functions of $u$ and $r$. Here $u$ is
the time--like coordinate: in the flat space--time $u$ is just the retarded
time, so that surfaces of constant $u$ represent null cones open to the
future; $r$ is a null coordinate ($g_{rr}=0$) such that surfaces
of constant $r$ and $u$ are spheres; $\theta$ and $\phi$ are the 
usual angular coordinates. Also, in this paper we use relativistic units
 ($G=c=1$).

The Einstein field equations, {\bf G} $=-8\pi${\bf T},
are considered with the energy--momentum tensor 
{\bf T} $=(\rho+p)${\bf v}$\otimes${\bf v}$-p${\bf g}$+${\bf q}$\otimes${\bf v}$+${\bf v}$\otimes${\bf q}$+${\bf E}, 
where {\bf v} and {\bf q} are the four--velocity and the heat
 flux four--vector, respectively, which must be orthogonal; {\bf E} is the
electromagnetic field energy--momentum tensor constructed with the Maxwell
field tensor {\bf F} as is usually done. The Maxwell field equations, 
{\bf d}$^*${\bf F}$=4\pi\,^*${\bf{J}} and {\bf dF}$=0$, 
are coupled minimally with gravitation, where {\bf J}$=\sigma${\bf v} is
the electric current four--vector.  

Because of the spherical symmetry, only the radial electric field
 $F^{ur}=-F^{ru}$ is 
 non--vanishing. If we define the function $C(u,r)$ by
the relation
\begin{equation}
F^{ur}=Ce^{-2\beta}/r^2,
\end{equation}
the inhomogeneous Maxwell equations 
become
\begin{equation}
C_{,r}=4\pi r^2 J^u e^{2\beta}
\end{equation}
and
\begin{equation}
C_{,u}=-4\pi r^2 J^r e^{2\beta},
\end{equation}
where the comma subscript represents partial derivative with respect to
 the indicated coordinate.
The function $C(u,r)$ is naturally interpreted as the charge within the
radius $r$ at time $u$. Thus, the inhomogeneous field equations
entail the conservation of
charge inside a sphere comoving with the fluid, expressed by
\begin{equation}
{\bf v}(C)=0.\label{cc}
\end{equation}

Let $\omega$ be the velocity of matter as seen by a
Minkowskian observer moving at $-\omega$ with respect to the local comoving
frame; the matter velocity in radiation coordinates is then given by
\begin{equation}
\label{mv}\frac{dr}{du}=\frac Vr\frac \omega {1-\omega }. 
\end{equation}
Introducing the mass function by
\begin{equation}
\label{mf}V=e^{2\beta}(r-2\tilde m(u,r)+C^2/r), 
\end{equation}
we can write the Einstein equations as
\begin{equation}
\label{ec1}
\frac{\rho +p\omega ^2}{1-\omega ^2}+\frac{2\omega q}{1-\omega ^2}= 
\frac{e^{-2\beta}(C C_{,u}/r-\tilde m_{,u})}{4\pi r(r-2\tilde m+{C^2}/r)}
+\frac {\tilde m_{,r}-C C_{,r}/r}{4\pi r^2}, 
\end{equation}

\begin{equation}
\label{ec2}\frac{\rho -p\omega }{1+\omega }-\left( \frac{1-\omega }{1+\omega 
}\right) q=\frac{\tilde m_{,r}-CC_{,r}/r}{4\pi r^2}, 
\end{equation}

\begin{equation}
\label{ec3}\left(\frac{1-\omega}{1+\omega}\right)(\rho +p)-2\left(\frac{
1-\omega}{1+\omega}\right)q=\frac {\beta_{,r}}{2\pi r^2}(r-2\tilde m+ 
{C^2}/r), 
\end{equation}
$$
p=-\frac{1}{4\pi}\beta_{,ur}\,e^{-2\beta}+\frac 1{8\pi }(1-
2\tilde m/r+C^2/r^2)(2\beta_{,rr}+4\beta_{,r}^2-
\beta_{,r}/r) 
$$
$$
+\frac 1{8\pi r}\left[3\beta_{,r}(1-2\tilde m_{,r})-\tilde m_{,rr}\right]
+ \frac{3\beta_{,r}}{8\pi r}(2CC_{,r}/r-C^2/r^2) 
$$
\begin{equation}
\label{ec4}+\frac 1{8\pi r^2}(C_{,r}^2+CC_{,rr}-2CC_{,r}/r).
\end{equation}

We describe the exterior space--time by the Reissner--Nordstr\"om--Vaidya
metric 
\begin{equation}
\label{em}ds_{+}^2=\left( 1-\frac{2m(u)}r+\frac{C_T^2}{r^2}\right)
du^2+2du\,dr-r^2\left( d\theta ^2+\sin {}^2\theta \, d\phi ^2\right), 
\end{equation}
where $m(u)$ is the total mass and $C_T$ the total charge.
It can be shown that the junction conditions to match the metrics (\ref{im})
and (\ref{em}), across the moving boundary surface $r=a(u)$, are
equivalent to the continuity of the functions $\tilde m(u,r)$ and $\beta (u,r)$
across the boundary, that is, $\tilde m(u,a)=m(u)$ and $\beta (u,a)=0$,
and to the equation 
\begin{equation}
\label{csf}-\beta _{,ua}+\bigg(1-\frac{2m}{a}+\frac{C_T^2}{a^2}\bigg)
\beta_{,ra}
-\frac{\tilde m_{,ra}}{2a}+\frac{C_T C_{,ra}}{2a^2}=0, 
\end{equation}
where the subscript $a$ indicates that the quantity is evaluated at the
surface $r=a(u)$.
We have used the continuity of the radial electric field through the
boundary assuming no surface free charge density, yielding $C(u,a)=C_T$.
It should be mentioned that the discontinuity of the
pressure at the boundary, $p_a= q_a$,  
is a direct consequence of (\ref{mv}), (\ref{ec2}), (\ref{ec3}) and the
junction conditions \cite{bd96,ds87,bd89}.
\section{The surface equations}

One of the surface equations is
just (\ref{mv}) evaluated at $r=a$, which takes the form
\begin{equation}
\label{apch}\dot a=\left(1-\frac{2m}a+\frac{C_T^2}{a^2}\right)
\frac{\omega _a}{1-\omega _a%
}, 
\end{equation}
where (\ref{mf}) has been used as well as the junction conditions for \~m, $%
\beta $ and $C$. It is convenient to scale the variables by the initial mass 
$m(0)$, such that

$$
A\equiv \frac a{m(0)};\:M\equiv \frac m{m(0)};\:\frac u{m(0)}\rightarrow
u;\:\frac{C_T}{m(0)}\rightarrow C_T. 
$$
Defining the surface potential F and $\Omega $ by
\begin{equation}
F\equiv 1-\frac{2M}A+\frac{C_T^2}{A^2} 
\end{equation}
and
\begin{equation}
\label{omg}\Omega \equiv \frac 1{1-\omega _a}, 
\end{equation}
(\ref{apch}) can be written in the form
\begin{equation}
\label{ap}\dot A=F(\Omega -1), 
\end{equation}
which is the first surface equation.

The second surface equation relates the total mass loss rate to the energy
flux through the boundary surface. It has been shown 
\cite{bd96} that this equation can be written as
\begin{equation}
\label{fp}\dot F=\frac{2QF+(1-F)\dot A}A-\frac{C_T^2\dot A}{A^3}, 
\end{equation}
where 
\begin{equation}
Q=(2\Omega -1)(4\pi r^2 q)_{r=a}. 
\end{equation}
The third surface equation is the charge conservation law given by (\ref{cc}%
), evaluated at the boundary 
\begin{equation}
\label{contcar}C_{,ua}=-\dot AC_{,ra}. 
\end{equation}

The fourth surface equation is model dependent. This correspond to the
Bianchi identity $T_{r,\mu }^\mu =0$, given by
$$
\frac{\partial \tilde p}{\partial r}+\frac{\tilde \rho +\tilde p}
{(1-2\tilde m/r+C^2/r^2) }\left( 4\pi r\tilde p+\frac{%
\tilde m}{r^2}-\frac{C^2}{r^3}\right) -e^{-2\beta}\left( \frac{\tilde
\rho +\tilde p}{1-2\tilde m/r+C^2/r^2}\right)_{,u} 
$$
\begin{equation}
\label{hidro}=\frac 2r(p-\tilde p)+\frac{CC_{,r}}{4\pi r^4} 
\end{equation}
where $\tilde p=(p-\omega\rho)/(1+\omega)-(1-\omega)q/(1+\omega)$ and
$\tilde \rho=(\rho-\omega p)/(1+\omega)-(1-\omega)q/(1+\omega)$.
Equation (\ref{hidro}) is the generalization of that of 
Tolman--Oppenheimer--Volkov for non--static and charged radiative situations
\cite{bd96} (see Reference \cite{b71} for the static case). 
\section{Self--similar space--time}
Self-similarity is invariably defined by the existence of a homothetic
Killing vector field \cite{op90}. We shall assume that the spherical
distribution admits a one--parameter group of homothetic motions.
A homothetic vector field on the manifold is one that satisfies
$\pounds_\xi {g}=$2$n{g}$ on a local chart,
 where $n$ is a constant on the
manifold and $\pounds$ denotes the Lie derivative operator. If $n \ne 0$ we
have a proper homothetic vector field and it can always be scaled so as to have
$n = 1$; if $n = 0$ then $\xi$ is a Killing vector on the manifold
\cite{h88,cms94}. So, for a constant rescaling, $\xi$ satisfies 
\begin{equation}
\pounds_\xi{\bf g}=2{\bf g}
\end{equation}
and has the form
\begin{equation}
\xi =\Lambda (u,r)\partial_u  +\lambda (u,r)\partial_r.
\end{equation}

The homo\-thetic
 equations reduce to
\begin{equation}
\xi(X)=Z\xi(Z),\label{se1}
\end{equation}
\begin{equation}
\xi(Y)=0,\label{se2}
\end{equation}
where $\lambda=r$, $\Lambda=\Lambda(u)$, 
$X=\tilde m/r$,  $Y=\Lambda e^{2\beta}/r$ and $Z=C/r$.
Therefore, $X=X(\zeta)$, $Y=Y(\zeta)$ and $Z=Z(\zeta)$ are solutions
 if the self--similar
 variable is defined as
\begin{equation}
\zeta\equiv r\, e^{- \int du/\Lambda}.
\end{equation}
In the next section we propose a simple solution
which evolves toward staticity, rendering an inhomogeneous and dusty
fluid sphere.

\section{A simple example}
The simple solution $\tilde m=m\,r/a$,
$e^{2\beta}=r/a$ and $C=C_T\,r/a$ satisfies
the additional symmetry equations (\ref{se1}) and (\ref{se2}), the
charge conservation equation (\ref{contcar}), 
  the continuity of the radial electric field
and the continuity of the first 
fundamental form.
Now, with these solutions and the junction condition (\ref{csf})
(continuity of the second fundamental form),
 $\Omega$ is determined by
\begin{equation}
\Omega=\frac{1-F-C_T^2/A^2}{2F}.\label{omega}
\end{equation}

The heat flow at the surface is obtained from the conservation equation
(\ref{hidro}) evaluated at the surface, resulting in
\begin{equation}
Q=\frac{F-C_T^2/A^2}{2F}(1-2F-C_T^2/A^2).
\end{equation}
This last equation, together with (\ref{omega}), must be taken into
account to integrate numerically the equations (\ref{ap}) and (\ref{fp}). 
Using a standard Runge--Kutta algorithm
and the initial conditions $A(0)=3.50$ and $m(0)=1.00$, we study the
effect of charge on collapse.
Once the boundary evolution and its 
energetic are determined, we calculate the physical variables from the
field equations. Figures 1--7 show the results obtained for our simple 
solution. We shall discuss them in the next section.
\section{Discussion}
Figure 1 displays the evolution of the surface for
different values of the total charge. It is clear that the increase
of the total charge favors the collapse in a first stage of the 
evolution. Later the
collapse is halted, with damped oscillations rendering a distribution 
which is less compact for
  greater total charge. 
We show in Figure 2 how the
Bondi mass decreases and oscillates about the total charge (given positive)
until reaching the same value of $C_T$.
Figures 3--6 sketch the pressure $p$, the density $\rho$, the heat flow
$q$ and the matter velocity $dr/du$. Observe that the whole sphere of fluid
becomes dust ($p=0$ at all points) and inhomogeneous when staticity is reached.
It is interesting to note that the cooling proceeds with emission
 and absorption
of energy and consequently all the shells bounce and contract as a unit,
over and over, until reaching staticity in the whole sphere. Also it is
striking how the ratio $p/\rho$ is a function only of the Bondi time,
 i.e., it is the same at any point of the charged distribution (see Figure 7).

We would like to stress that our solution, in spite of its simplicity,
behaves very well when one takes into account the results
 reported by other authors.
The sphere collapses and rebounds as a unit \cite{l95}, over and over,
 until reaching
staticity.
Equilibrium configurations are possible, 
with no necessity of internal pressure; 
moreover, electric charge halts the gravitational collapse \cite{b65,b71}.
The spheres of charged matter can oscillate and the final static
configuration is reached when the total mass is equal to the total
charge \cite{f69}, that is, the charged spheres of dust in equilibrium 
belong to the interior Papapetrou--Majumdar class \cite{cd78}. 

Finally, we make special mention of the Herrera and Ponce de Le\'on
homothetic models \cite{hp85}.  For a null radial pressure at the 
surface (as a boundary condition) they find a charged dust. Otherwise,
the distribution is infinitely extended. From an altogether
 physical point of view,
perhaps it is possible to find finite perfect (and self--similar)
 fluid sources
with zero pressure at the surface, considering 
free--streaming as the inner transport mechamism. In general, Herrera
and Ponce de Le\'on find that the total charge is less than the
total mass. The equality will only be valid in the case of homothetic
charged dust spheres.

\section*{Acknowledgments}
We benefited from research support by the Consejo de Investigaci\'on
under Grant CI-5-1001-0774/96 of the Universidad de Oriente.
We would like to thank Loren Lockwood for his helpful reading and
valuable comments.

\newpage
\begin{figure}
\centerline{\epsfxsize=5in\epsfysize=5in\epsfbox{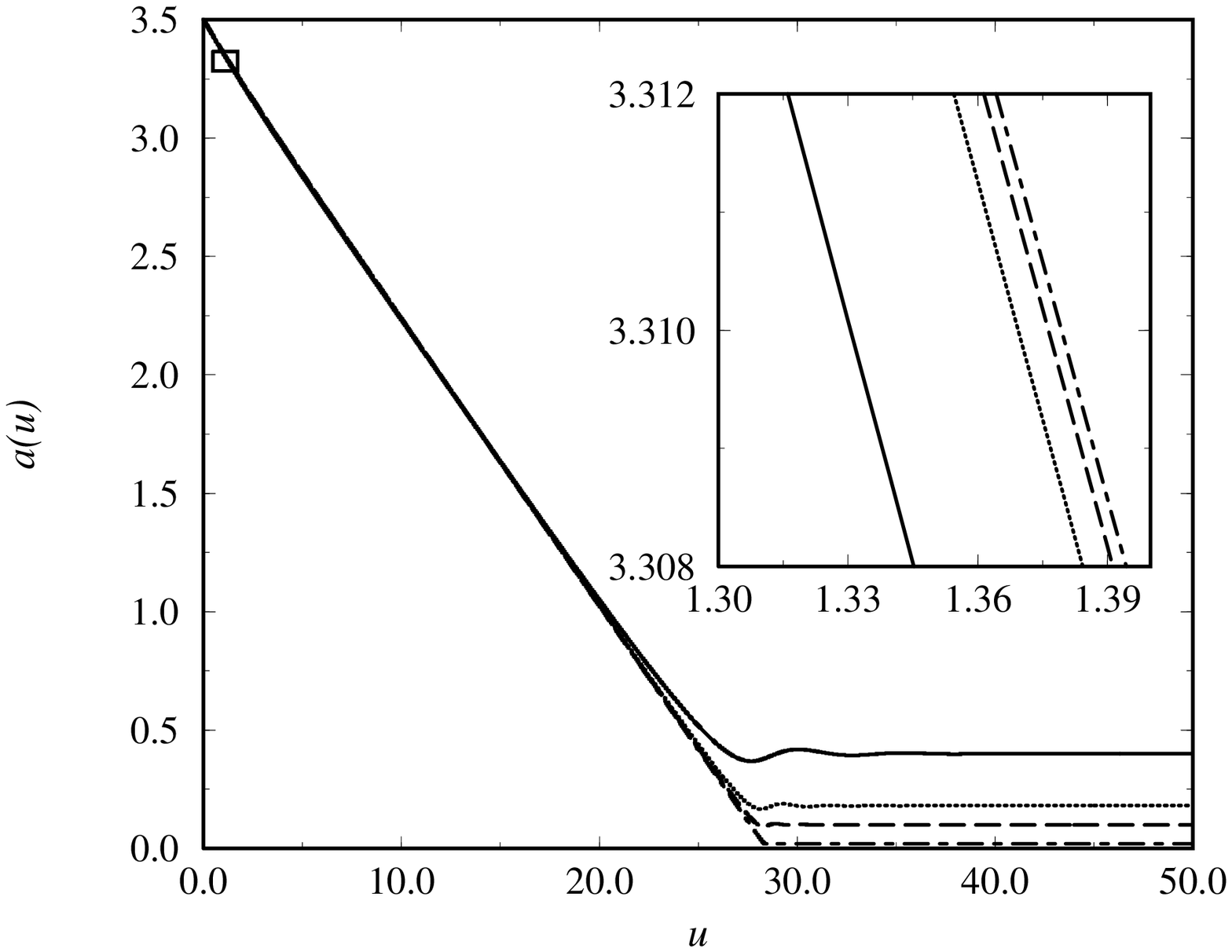}}
\caption{Radius $a$ as a function of the time $u$ for $A(0)=3.5$,
$m(0)=1$ and for different values of the
total charge $C_T$: $0.01$ (dot--dashed line); $0.05$ (dashed line);
 $0.09$ (dotted line); $0.2$ (solid line). 
}
\label{fig:radius}
\end{figure}
\begin{figure}
\centerline{\epsfxsize=5in\epsfysize=5in\epsfbox{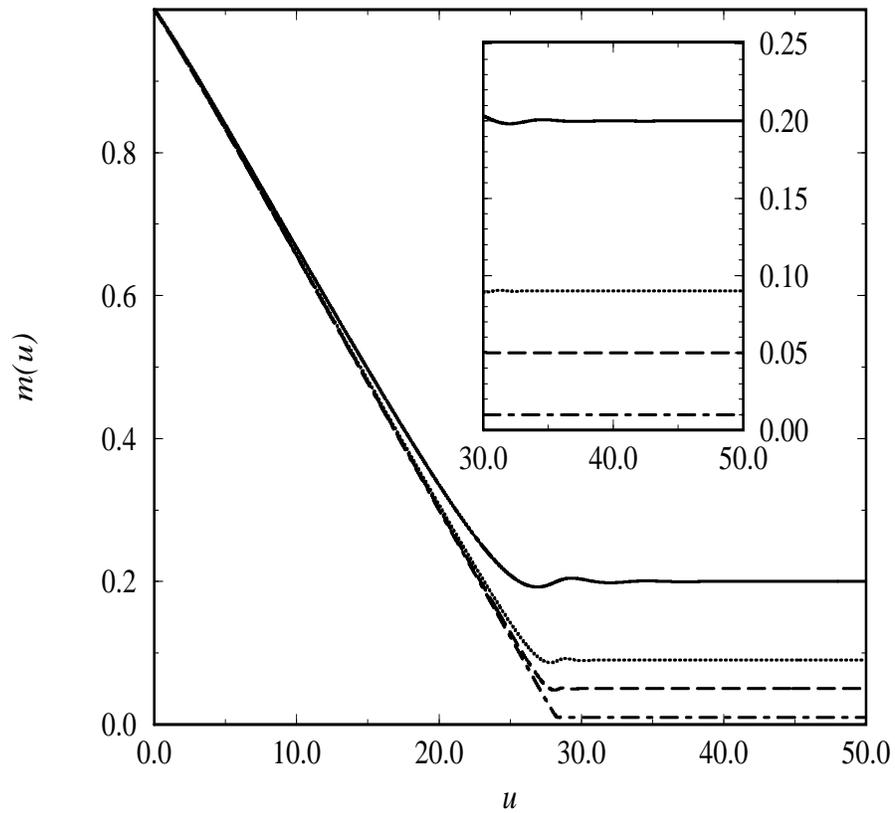}}
\caption{Bondi mass at the surface $m$ as a function of time $u$ for $A(0)=3.5$,
$m(0)=1$ and for different values of the
total charge $C_T$: $0.01$ (dot--dashed line); $0.05$ (dashed line);
 $0.09$ (dotted line); $0.2$ (solid line). Observe that the final
mass in each case is equal to the total charge.}
\label{fig:BM}
\end{figure}
\begin{figure}
\centerline{\epsfxsize=5in\epsfysize=5in\epsfbox{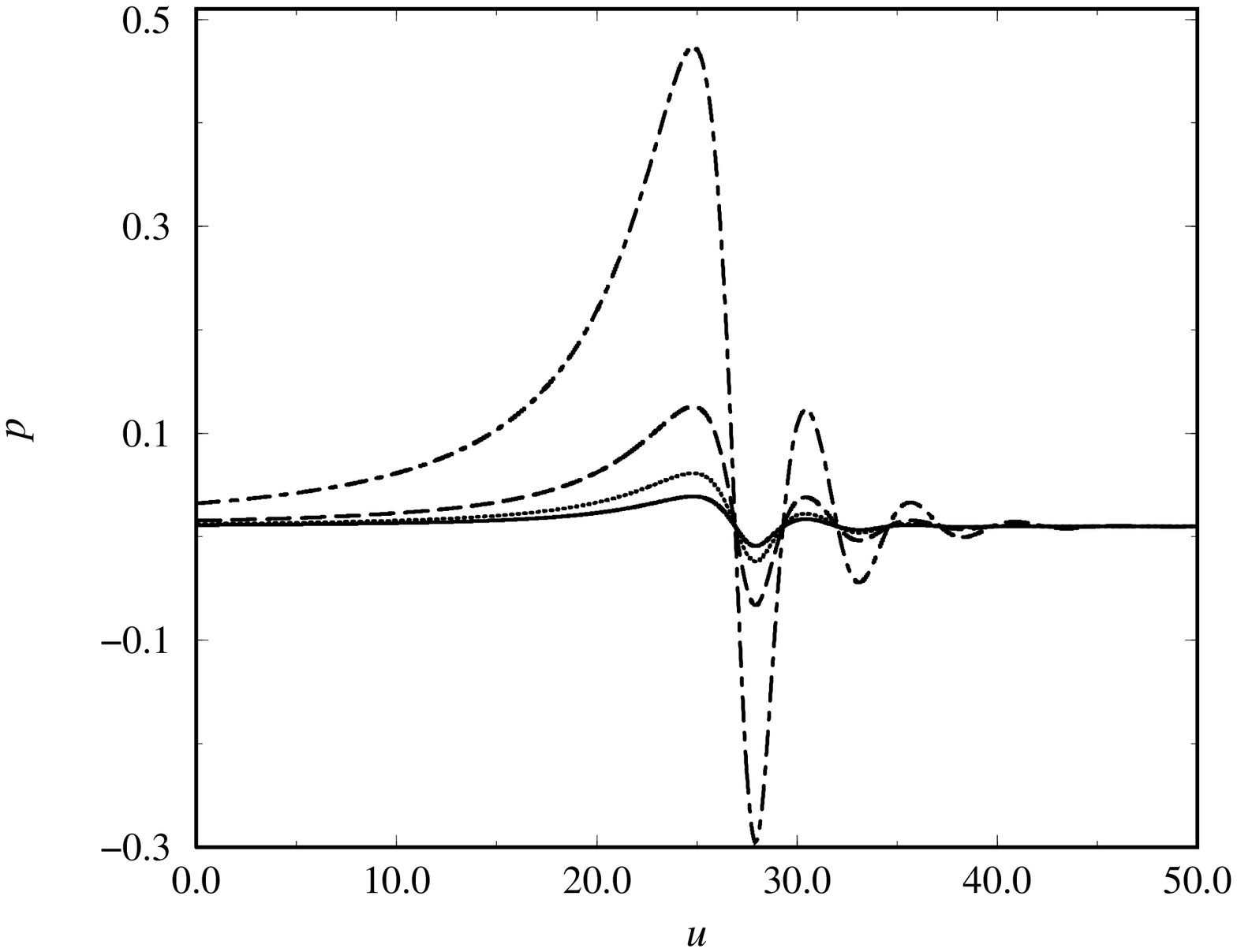}}
\caption{Pressure $p$ as a function of time $u$ for 
$A(0)=3.5$, $m(0)=1$ and $C_T=0.09$ at different points:
$r/a=0.25$ (dot--dashed line); $r/a=0.33$ (dashed line); $r/a=0.50$
(dotted line); r/a=1.00 (solid line).}
\label{fig:figure3}
\end{figure}
\begin{figure}
\centerline{\epsfxsize=5in\epsfysize=5in\epsfbox{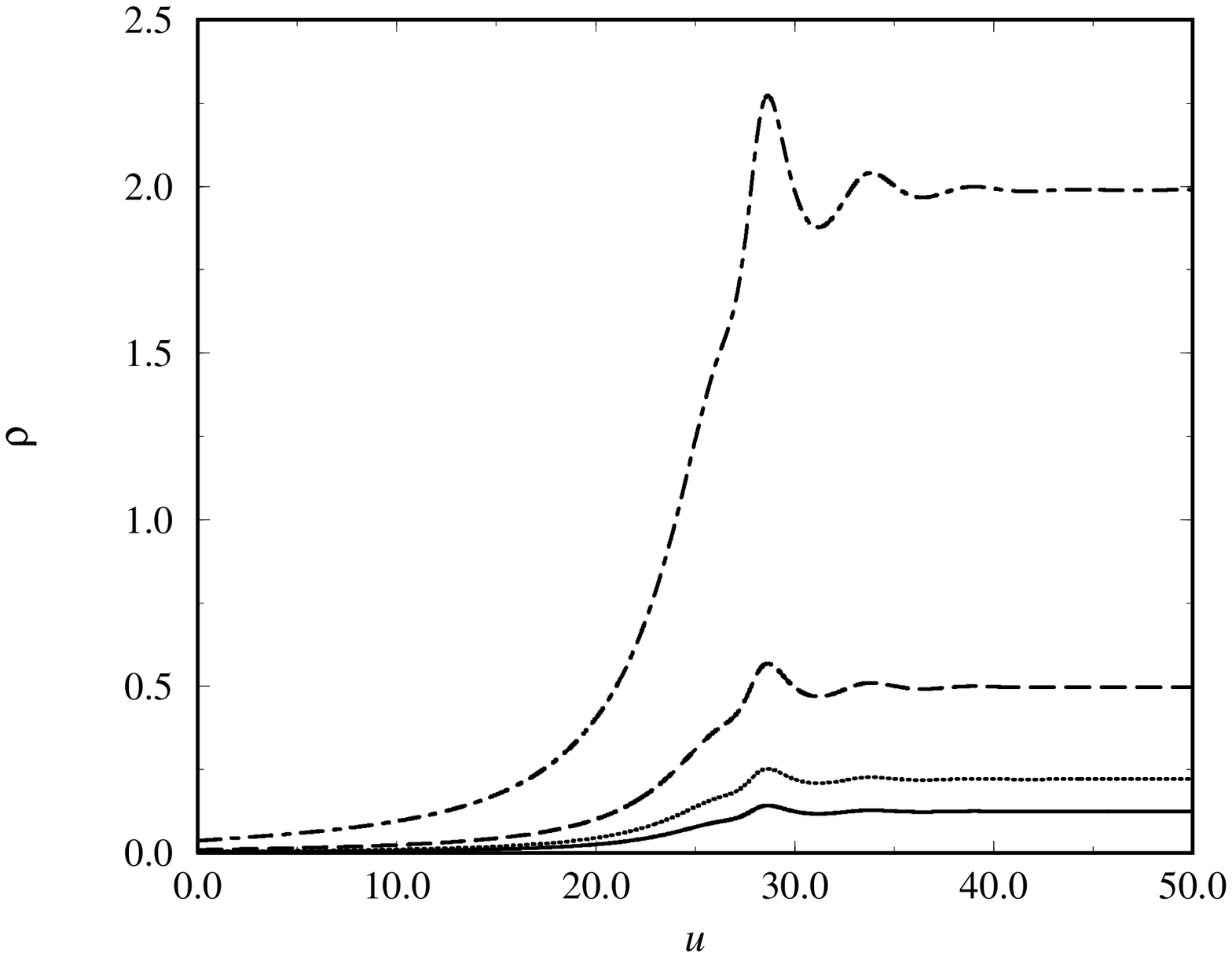}}
\caption{Density $\rho$ as a function of time $u$ for 
$A(0)=3.5$, $m(0)=1$ and $C_T=0.09$ at different points:
$r/a=0.25$ (dot--dashed line); $r/a=0.33$ (dashed line); $r/a=0.50$
(dotted line); r/a=1.00 (solid line).}
\label{fig:figure4}
\end{figure}
\begin{figure}
\centerline{\epsfxsize=5in\epsfysize=5in\epsfbox{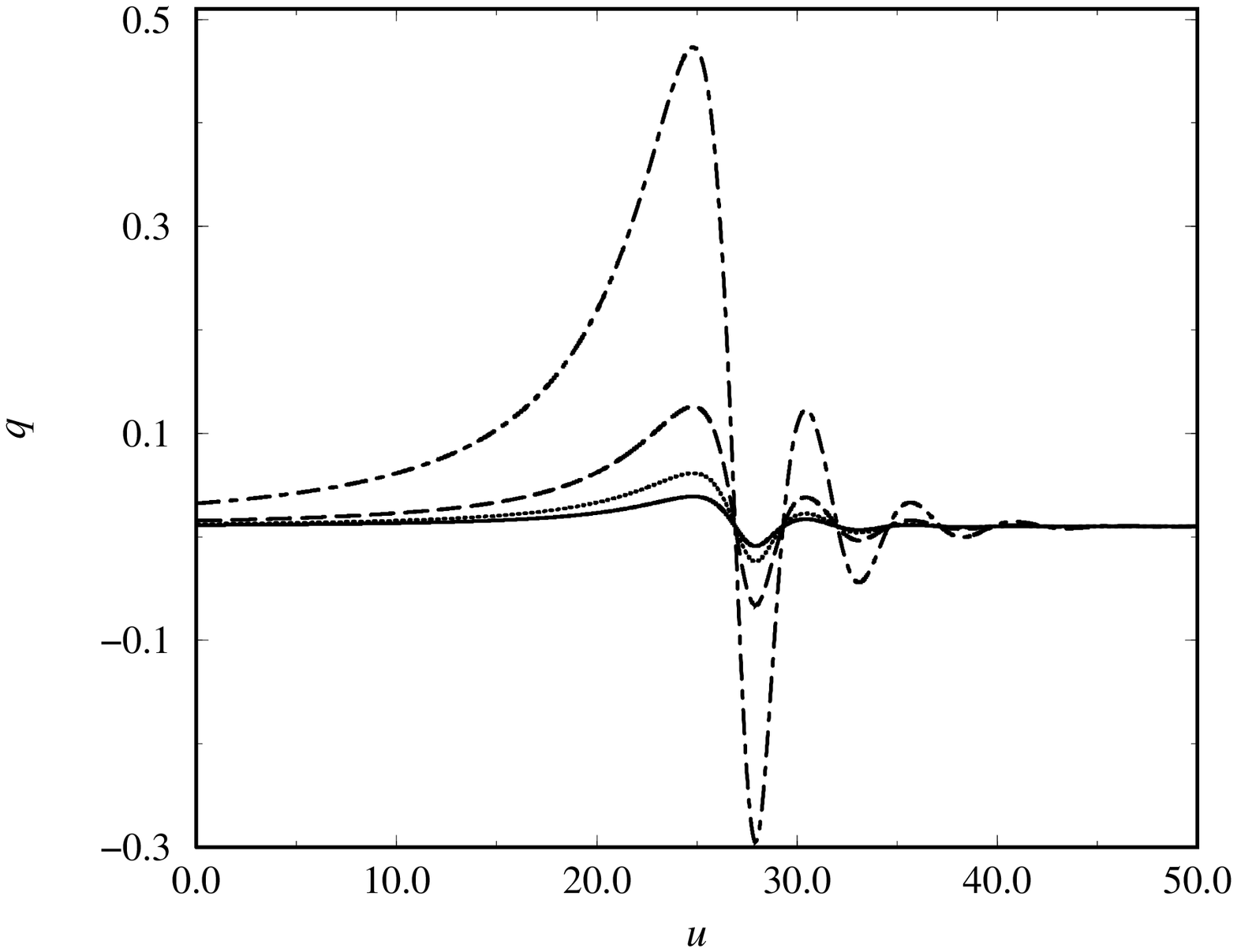}}
\caption{Heat flow $q$ as a function of time $u$ for 
$A(0)=3.5$, $m(0)=1$ and $C_T=0.09$ at different points:
$r/a=0.25$ (dot--dashed line); $r/a=0.33$ (dashed line); $r/a=0.50$
(dotted line); r/a=1.00 (solid line).}
\label{fig:figure5}
\end{figure}
\begin{figure}
\centerline{\epsfxsize=5in\epsfysize=5in\epsfbox{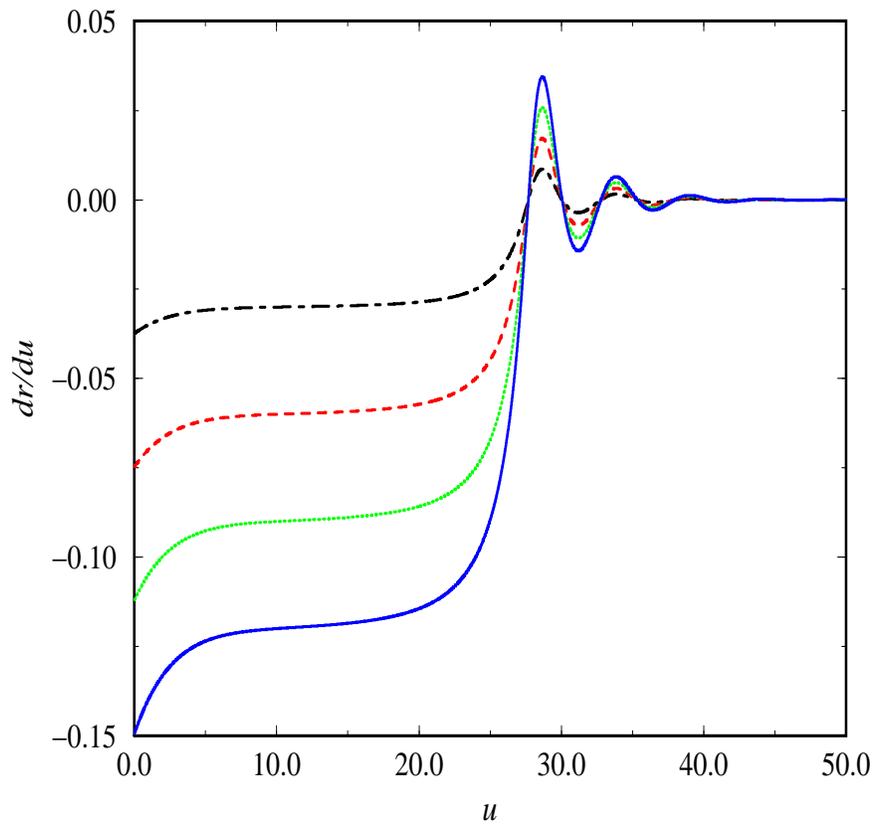}}
\caption{Matter velocity $dr/du$ as a function of time $u$ for 
$A(0)=3.5$, $m(0)=1$ and $C_T=0.09$ at different points:
$r/a=0.25$ (dot--dashed line); $r/a=0.33$ (dashed line); $r/a=0.50$
(dotted line); r/a=1.00 (solid line).}
\label{fig:figure6}
\end{figure}
\begin{figure}
\centerline{\epsfxsize=5in\epsfysize=5in\epsfbox{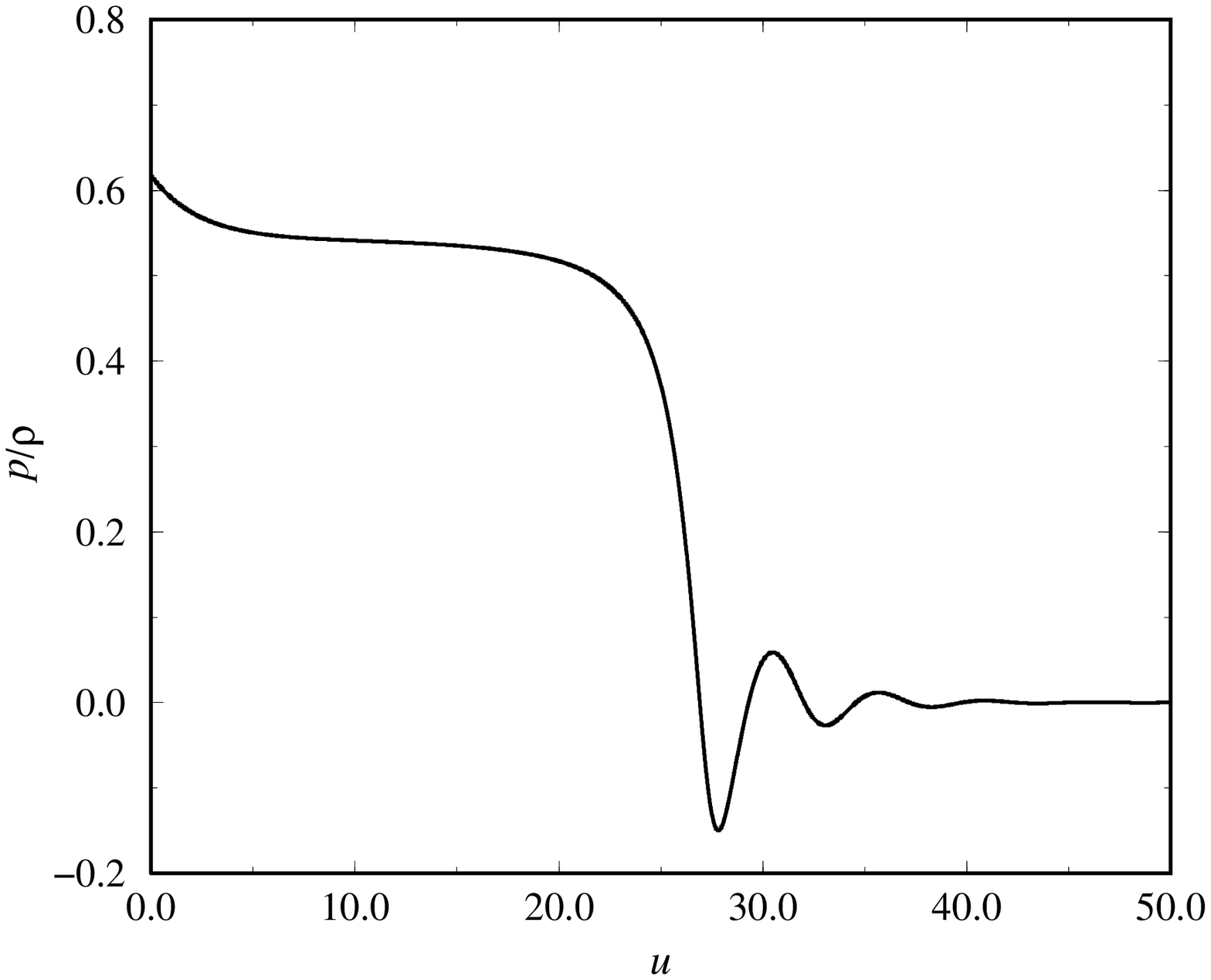}}
\caption{Ratio $p/\rho$ as a function of time $u$ for
$A(0)=3.5$, $m(0)=1$ and $C_T=0.09$ at all points.}
\label{fig:figure7}
\end{figure}

\section*{References}
\begin{thebibliography}{99}
\bibitem{b65} Bonnor W B 1965 {\it Mon. Not. Royal Astr. Soc.} {\bf 129} 443
\bibitem{f69} Faulkes M C 1969 {\it Can. J. Phys.} {\bf 47} 1989
\bibitem{s71} Shvarsman V F 1971 {\it Sov. Phys. JETP} {\bf} 33 475
\bibitem{b71} Bekenstein J D 1971 {\it Phys. Rev. D}{\bf 4} 2185
\bibitem{b75} Bonnor W B 1975 {\it Mon. Not. Royal Astr. Soc.} {\bf 170} 643
\bibitem{cd78} Cooperstock F I and de la Cruz V 1978 {\it Gen. Rel. Grav.} {\bf 9} 835
\bibitem{hp85} Herrera L and Ponce de Le\'on J 1985 {\it J. Math. Phys.} {\bf 26} 2302
\bibitem{o90} Ori A 1990 {\it Class. Quantum Grav.} {\bf 7} 985
\bibitem{lz91} Lake K and Zannias T 1991 {\it Phys. Rev. D } {\bf 43} 1798
\bibitem{o91} Ori A 1991 {\it Phys. Rev. D} {\bf 44} 2278
\bibitem{fms95} Fayos F, Martin-Prats M M and Senovilla J M M 1995 {\it Class.
Quantum Grav.} {\bf 12} 2565
\bibitem{p93} Ponce de Le\'on J  1993 {\it Gen. Rel. Grav} {\bf 25} 1123
\bibitem{l95} L\'opez C 1995 {\it Gen. Rel. Grav.} {\bf 27} 85
\bibitem{ks79} Kazanas D and Schramm D 1979 {\it Sources of gravitational 
Radiation} (Cambridge: Cambridge University Press)
\bibitem{ds87} de Oliveira A K G and Santos N O  1987 {\it Ap. J} {\bf 312} 640
\bibitem{bd89} Banerjee A and Dutta Choudhury S B 1989 {\it Gen. Rel. Grav.}
 {\bf 21} 785
\bibitem{ct71} Cahill M E and Taub A H 1971 {\it Commun. Math. Phys. } {\bf 21} 1
\bibitem{op90} Ori A and Piran T 1990 {\it Phys. Rev. D} {\bf 42} 1068
\bibitem{hp91} Henriksen R Patel K 1991 {\it Gen. Rel. Grav.} {\bf 23} 527
\bibitem{lz90} Lake K and Zannias T 1990 {\it Phys. Rev. D } {\bf 41} 3866
\bibitem{ch74} Carr J and Hawking S 1974 {\it Mon. Not. Royal Astr. Soc.} {\bf 168} 399
\bibitem{bh78a} Bicknell G V and Henriksen N R 1978 {\it Ap. J} {\bf 219} 1043
\bibitem{bh78b} Bicknell G V and Henriksen N R 1978 {\it Ap. J.} {\bf 225} 237
\bibitem{eimm86} Eardley D M, Isemberg J, Mardsden J and Moncrief V 1986 {\it
Comm. Math Phys.} {\bf 106} 137
\bibitem{cy90} Carr J and Yahil A 1990 {\it Ap. J} {\bf 360} 330
\bibitem{l92} Lake K 1992 {\it Phys. Rev Lett.} {\bf 68} 3129
\bibitem{b95} Brady P R 1995 {\it Phys. Rev. D} {\bf 51} 4198
\bibitem{cc98} Carr B J and Coley A A 1998 {\it gr-qc/9806048}
\bibitem{ps79} Pant D N and Sah A 1979 {\it J. Math. Phys.} {\bf 20} 2537
\bibitem{t84} Tikekar R 1984 {\it J. Math. Phys.} {\bf 25} 1481
\bibitem{hm84} Humi M and Mansour J 1984 {\it Phys. Rev. D} {\bf 29} 1076
\bibitem{b64} Bondi H 1964 {\it Proc. R. Soc. London A} {\bf 281} 39
\bibitem{bd96}  Barreto W and Da Silva A 1996 {\it Gen. Rel. Grav.} {\bf 28}
735
\bibitem{h88} Hall G S 1988 {\it Gen. Rel. Grav.} {\bf 20} 671
\bibitem{cms94} Carot J, Mas L and Sintes A M 1994 {J. Math. Phys.} {\bf
35} 3560 
 
\end{thebibliography}
\end{document}